\documentclass{article}

\usepackage{arxiv}

\usepackage[utf8]{inputenc} 
\usepackage[T1]{fontenc}    
\usepackage{hyperref}       
\usepackage{url}            
\usepackage{booktabs}       
\usepackage{amsfonts}       
\usepackage{nicefrac}       
\usepackage{microtype}      
\usepackage{lipsum}		
\usepackage{} 
\usepackage{graphicx}
\usepackage{natbib}
\usepackage{doi}
\usepackage{amssymb}
\usepackage{url,hyperref,lineno,microtype,subcaption}
\usepackage[onehalfspacing]{setspace}
\usepackage{graphicx}
\usepackage{CJKutf8}
\usepackage{amsmath}
\usepackage{dirtytalk}
\usepackage{algorithm}
\usepackage{varwidth}
\usepackage{mathtools}
\usepackage{multirow}
\usepackage{siunitx}
\usepackage[utf8]{inputenc}
\usepackage{tabto}
\usepackage{algpseudocode} 
\newcounter{algsubstate}
\renewcommand{\thealgsubstate}{\alph{algsubstate}}
\newenvironment{algsubstates}
  {\setcounter{algsubstate}{0}%
   \renewcommand{\State}{%
     \stepcounter{algsubstate}%
     \Statex {\footnotesize\thealgsubstate:}\space}}
  {}

\title{A Multi-Opinion Based Method for Quantifying Polarization on Social Networks}

\author{Maneet Singh \\
	Department of Computer Science and Engineering \\
	Indian Institute of Technology, Ropar, India \\
	\texttt{2018csz0008@iitrpr.ac.in} \\
	\And
	S.R.S. Iyengar\\
	Department of Computer Science and Engineering \\
	Indian Institute of Technology, Ropar, India \\
	\texttt{sudarshan@iitrpr.ac.in}
	\And Rishemjit Kaur\\
	CSIR-Central Scientific Instruments Organisation, Chandigarh, India \\
	Academy of Scientific and Innovative Research, Ghaziabad, India \\
	\texttt{rishemjit.kaur@csio.res.in} \\
}

\date{}

\renewcommand{\headeright}{}
\renewcommand{\undertitle}{}
\renewcommand{\shorttitle}{\textit{A Multi-Opinion Based Method for Quantifying Polarization on Social Networks} }

\hypersetup{
pdftitle={A template for the arxiv style},
pdfsubject={q-bio.NC, q-bio.QM},
pdfauthor={David S.~Hippocampus, Elias D.~Striatum},
pdfkeywords={First keyword, Second keyword, More},
}

\begin{document}
\maketitle

\begin{abstract}
Social media platforms have emerged as a hub for political and social interactions, and analyzing the polarization of opinions has been gaining attention. In this study, we have proposed a measure to quantify polarization on social networks. The proposed metric, unlike state-of-the-art methods, does not assume a two-opinion case and applies to multiple opinions. We tested our metric on different networks with a multi-opinion scenario and varying degrees of polarization. The scores obtained from the proposed metric were comparable to state-of-the-art methods on binary opinion-based benchmark networks. The technique also differentiated among networks with different levels of polarization in a multi-opinion scenario. We also quantified polarization in a retweet network obtained from Twitter regarding the usage of drugs like hydroxychloroquine or chloroquine in treating COVID-19. Our metric indicated a high level of polarized opinions among the users. These findings suggest uncertainty among users in the benefits of using hydroxychloroquine and chloroquine drugs to treat COVID-19 patients.
\end{abstract}


\maketitle

\section{Introduction}
The existence of liberals and conservatives in the European countries or the division of Americans into Democrats and Republicans are examples of disagreements along the lines of political ideologies. Such disagreements in public discourse for any given issue leads to opinion polarization. The term \say{opinion polarization} refers to the segregation of individuals based on their opinion towards a given issue. The existence of polarization has been observed among users engaged in discussing and expressing their opinion towards a given topic \citep{valenzuela2013unpacking} on online platforms like  Twitter, Facebook, etc. \citep{conover2011political}. 
There has been various studies that analyses the presence of polarized opinion for variety of issues such as EU referendum of UK or Brexit referendum \citep{bossetta2018political}, the compulsion of licence for carrying a gun  \citep{miller2019americans}, right to abortion \citep{mouw2001culture}, laws related to LGBT \citep{lewis2017degrees}, historical \citep{goldberg2018teaching}, racial \citep{gallagher2018divergent} and vaccination \citep{yuan2019examining}.

Besides qualitative assessment of the polarized opinions on social networks of users engaged in the discussions of any given issue, the prior research has also focused on quantitative assessment of opinion polarization. The measurement of polarization among users can be broadly classified into two categories based on the studies conducted: The first is based primarily on user opinion \citep{dimaggio1996have,gay1996search,banisch2019opinion, koudenburg2021new}, whereas the second is based solely on the structure of a social network of users \citep{morales2015measuring,garimella2018political}. The first set of studies examines polarization from various perspectives, including dispersion, i.e., the degree of divergence or distance between opposing opinions, and bi-modality, which employs the concept of bimodal distribution and thus focuses on the formation of two opinion clusters regardless of the distance between them \citep{dimaggio1996have,gay1996search,banisch2019opinion, koudenburg2021new}. These methods do not use the network structures among the users that could provide additional information about the social relationships and other homophilic preferences. 

The second line of research is based on the notion that the polarization of social networks is a proxy for polarized opinions and so measures opinion polarization purely based on network characteristics. Based on the above assumption, modularity \citep{newman2006modularity} has been correlated with polarization. Here, a network with high modularity indicates dense connections among the users belonging to the same community (or module) and similarly sparse connections among users belonging to different communities. Later,  \citep{guerra2013measure} argued that the number of high-degree users present at the boundaries of two potentially polarized communities is a better estimate than modularity for quantifying polarization for issues like gun control. Another study \citep{morales2015measuring} introduced an electric dipole moment-based opinion polarization metric for quantifying the level of polarization in Twitter discussions related to the late president of Venezuela. The method uses the opinion probability distribution function, obtained from the given social network. Later, Garimella et al. \citep{garimella2017long} proposed various measures for quantifying polarization for controversial and non-controversial issues. These methods were applied on retweet networks and follower networks of users engaged in the discussions of those issues. These measures involve using different network properties such as random walks, graph embeddings, betweenness centrality, and boundary connectivity \citep{guerra2013measure,garimella2018quantifying}. The methods, such as random walks and user embeddings, were later modified to make them more efficient \citep{darwish2019quantifying}. 

As mentioned previously, the existing network-based methods of quantifying opinion polarization assume network polarization to be equivalent to opinion polarization. These methods generally divide the network into two communities, each representing an opinion group. Here a community within a network corresponds to a set of users with strong connections among them \cite {girvan2002community,radicchi2004defining}. The above community-based opinion assignment approach assumes that users within a community possess the same opinions, whereas there are different opinions across the communities. But a community may have users with different opinions \citep{yuan2019examining}. Similarly, a given network may contain more than two communities\citep{labatut2014identifying,yuan2019examining}, and hence there may be more than one community belonging to a group of people having a similar opinion. Another issue with forming two opinion groups from the network structure is considering opinion to be binary. This assumption ignores that there might be more than two opinions towards a given issue, for instance, the existence of a neutral opinion \citep{singh2022mining} or multi-sided controversial issues \citep{ani_news2019}. Hence, it becomes vital to quantify polarization with respect to communication between users with different opinions belonging to similar or different communities. To the best of our knowledge, there has not been any study that quantifies polarization on a social network with the multi-opinion scenario.
 
The method proposed in our study provides a polarization score for a given network with the multi-opinion scenario by combining the scores obtained using two disjoint sets of edges within the network. The first score focuses only on quantifying polarization by considering edges connecting users of the same community. On the other hand, the second score mainly focuses on quantifying polarization across the communities by considering edges connecting users belonging to different communities. The performance of the proposed metric has been verified by first applying it on benchmark networks with binary opinions to compare it with the state-of-the-art methods. To further demonstrate the feasibility of our proposed method, we implemented it on synthetic networks with a different number of distinct opinions among the users and having varying degrees of polarization. We also employed our multi-opinion metric for quantifying polarization related to Twitter discussions regarding the prevention or treatment of COVID-19 using hydroxychloroquine and chloroquine drugs \citep{mutlu2020stance}. These drugs have anti-inflammatory and anti-viral properties \citep{sinha2020hydroxychloroquine} due to which their emergency usage was approved by US FDA for patients suffering from COVID-19 \citep{piszczatoski2020emergency}. The approval was based on only preliminary analysis as there is no clear evidence that usage of the drug is helpful for COVID-19 patients \citep{li2020hydroxychloroquine}. Due to this, the approval was later revoked by US FDA \citep{manivannan2021rise}. We have used the discussions regarding these drugs to quantify polarization that might have occurred because of the uncertainty associated with using the drugs for the treatment of COVID-19 patients. Thus, the two key contributions of our work can be stated as follows:
\begin{enumerate}
    \item Proposing a multi-opinion-based polarization quantification method and testing its efficiency on networks with different polarization levels and varying numbers of opinions among users.
    \item Using the proposed approach for quantifying polarization in a tri-opinion (\say{favor}, \say{against} and \say{neutral}) based network representing retweeting behavior towards the given issue among users.

\end{enumerate}

\section{Methodology}
In this section, we have first described our proposed multi-opinion based metric for quantification of polarization. Next, we have introduced the benchmark networks with binary opinion. We have also described the construction mechanism of synthetic networks used for testing the efficiency of the proposed method. Finally, we constructed a tri-opinion-based real social network, which includes steps like collecting the tweets on a given issue, extracting details of users retweeting those tweets, and finally constructing a retweet network of users of all the tweets.

\subsection{Proposed Polarization Metric}

The paper proposes a multi-opinion-based approach for measuring the level of polarization between the users with different opinions in any given network \footnote[1]{Code and data for the paper is available at \href{https://figshare.com/s/daac579eee2a60dfcbe9}{https://figshare.com/s/daac579eee2a60dfcbe9}}. It utilizes the users' opinions and the network's community structure. The proposed methodology divides the edges between users in the given network into two categories: within-community edges and between-community edges. Based on this division, we computed two different scores, where both focus on different sets of edges in the network. The first corresponds to the within-community polarization score ($P_W$) used for measuring polarization among users within the same community and hence uses the edges connecting users belonging to the same communities in the network. Similarly, the second score ($P_B$) captures the polarization between users belonging to different communities. The final score was then computed using the weighted average of these two scores. The complete approach is described in detail below:-


\begin{figure}
  \centering
  \includegraphics[width=\linewidth]{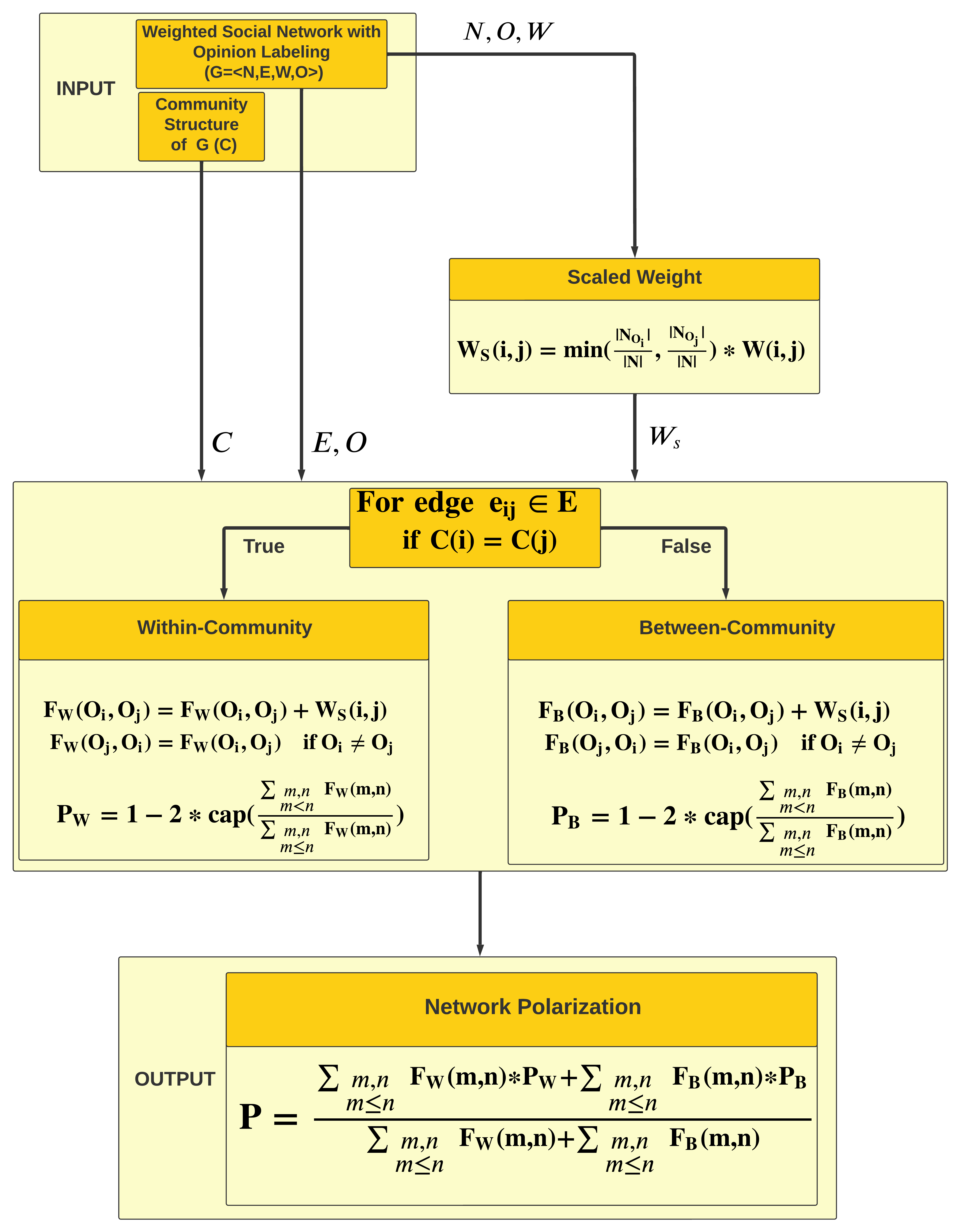}
  \caption{Block Based Diagrammatic Representation of Proposed Multi-Opinion Based Polarization Metric. The input to the method is a weighted social network $G$ with the opinion information of all the users and the community structure of $G$. The final output is the polarization score for the network $G$.}
 
  \label{fig:blocks}
\end{figure}

\begin{algorithm}[H]
	\caption{Quantifying Polarization using Multi-Opinion Based Methodology} 
	 \begin{flushleft}
    \textbf{Input} A weighted network $G<N,E,W,O>$ and the set of communities $C$ within the network $G$ (obtained using a community detection algorithm). \\
    
    \textbf{Output} Polarization Score of the network 
    \end{flushleft}
	\begin{algorithmic}[1]
		
		\State 	Let $N$ be the set of users, $E$  be the set of edges and $W$ be the weight matrix (of size $|N|*|N|$) for the network $G$.

		\State Compute the scaled version of the weights $W$ for each edge $e(i,j)$ (to give more weightage to opinions possessed by majority of users) as follows:-
		\begin{equation}
		\label{eq:1}
		    W_{S}(i,j)=avg(\frac{|N_{O_i}|}{|N|},\frac{|N_{O_j}|}{|N|})*W(i,j)
		\end{equation}
		Here $avg$ is the function to compute average of two numbers, $O_i$ and $O_j$ are the opinions of user $i$ and $j$ respectively and $N_{O_i}$ and $N_{O_j}$ represents the set of users with opinion $O_i$ and $O_j$ respectively.
		\State 	Let $F_{W}$ be the within-community frequency matrix of size $numOpinions * numOpinions$ (initially set to contain all zeros) for storing the number of weighted edges connecting users belonging to same community for all possible pairs of opinions. Similarly between-community frequency matrix $F_{B}$ of size $numOpinions * numOpinions$ (initially set to contain all zeros) is used for edges connecting users belonging to different communities.
		\State 	For each edge $e(i,j)$ (connecting user $i$ and $j$) in $E$:-
		\begin{algsubstates}
		\State 	If $C(i)=C(j)$ then update  $F_W$ as follows:- 
		\begin{gather}
		\label{eq:2}
		    F_{W}(O_i,O_j)=F_{W}(O_i,O_j)+W_{c}(i,j)\\
		    F_{W}(O_j,O_i)=F_{W}(O_i,O_j) \, \, if\,\, O_i \neq O_j
		\end{gather}
		\State 	If $C(i)\neq C(j)$ then update  $F_B$ as follows:-
		\begin{gather}
		\label{eq:4}
		    F_{B}(O_i,O_j)=F_{B}(O_i,O_j)+W_{c}(i,j)\\
		    F_{B}(O_j,O_i)=F_{B}(O_i,O_j) \, \, if\,\, O_i \neq O_j
		\end{gather}
		\end{algsubstates}
		\State 	The within-community polarization score ($P_W$) is computed as follows:-
		\begin{equation}
		\label{eq:6}
            P_W=1-2*cap(\frac{\sum_{\substack{m,n \\ m<n}} F_{W}(m,n)}{\sum_{\substack{m,n \\ m\leq n}} F_{W}(m,n)}) 
        \end{equation}
        where \begin{equation*}
            cap(X)=
            \begin{cases}
                X & \text{if } X<0.5\\
                0.5 & otherwise
            \end{cases}
        \end{equation*}
		\State The between-community polarization score ($P_B$) is computed as follows:-
		\begin{equation}
		\label{eq:7}
            P_B=1-2*cap(\frac{\sum_{\substack{m,n \\ m<n}} F_{B}(m,n)}{\sum_{\substack{m,n \\ m\leq n}} F_{B}(m,n)}) 
        \end{equation}
    \State 	Using $P_W$ and $P_B$, the final opinion based network polarization score ($P$) can be obtained as:-
    \begin{equation}
    \label{eq:8}
            P=\frac{\sum_{\substack{m,n \\ m\leq n}} F_{W}(m,n)*P_{W}+\sum_{\substack{m,n \\ m\leq n}} F_{B}(m,n)*P_{B}}{\sum_{\substack{m,n \\ m\leq n}} F_{W}(m,n)+\sum_{\substack{m,n \\ m\leq n}} F_{B}(m,n)} 
        \end{equation}
        	\end{algorithmic}
	\label{alg:polarization}
\end{algorithm}

The proposed polarization quantification method requires two parameters as input. First is a weighted social network $G=<N, E, W, O>$. Here, $N$ corresponds to the set of users, $E$ represents a set of edges connecting the users depending upon the network (for instance, in a retweet network, an edge between user A and B would represent that either A's tweet is retweeted by B or B's tweet is retweeted by A \citep{conover2011political}). The weight matrix $W$ in the network $G$ indicates the strength of connections ($E$) between the users. For instance, in a retweet network, the weight could be the sum of the number of tweets of A retweeted by B and the number of tweets of B retweeted by A. In the network $G$, each user is assumed to be labeled with their opinion ($O$) towards any given issue. The opinion in our case can take on any discrete number of values ($numOpinions$), which can be greater than or equal to two. In other words, the algorithm applies to scenarios involving different levels of opinion (such as 'pro', 'anti', and \say{Neutral}) or multi-sided controversies. The second input to the algorithm is the set of communities ($C$) within the network ($G$), obtained using a Louvain community detection algorithm \citep{blondel2008fast}.

For computing the opinion-based polarization score for a given network, we first scaled the weights $W$ for each edge $e(i,j)$ based on the average of the fraction of users possessing the opinions $O_i$ and $O_j$ (Eq \ref{eq:1}). The above factor ensures that if any opinion $O_k$ is possessed by a small fraction of users, then an edge having user(s) with opinion $O_k$ would contribute relatively less to the overall polarization score. The motivation behind this scaling is from the definition of polarization given in the literature that, besides segregation, also considers the size of each opinion group for assessing or computing the score of polarization \citep{morales2015measuring,olivares2019opinion}. Next, we compute the weighted sum of all the edges connecting users belonging to same community (i.e. within-frequency matrix $F_W$ of size $|numOpinions|*|numOpinions|$) using Eq \ref{eq:2}. Similarly, we also computed  the weighted sum of all the edges connecting users belonging to different communities (i.e. between-frequency matrix $F_B$ of size $|numOpinions|*|numOpinions|$) using Eq \ref{eq:4}. Eq 3 and 5 compute the remaining entries of within-frequency $F_W$, and between-frequency matrices as the number of weighted edges for opinion $O_i$ and $O_j$ are similar to the number of weighted edges for opinion $O_j$ and $O_i$ for an undirected network. Later, we computed the within-community score ($P_W$) using $F_W$ (Eq \ref{eq:6}) and the between-community score ($P_B$) using $F_B$ (Eq \ref{eq:7}). The key intuition for Eq \ref{eq:6} and \ref{eq:7} is finding the weighted proportion of edges connecting users with different opinions from the given edges for each score. A function $cap$ in Eq \ref{eq:6} and \ref{eq:7} is used to ensure that the score will always be zero if the number of weighted edges between opposing individuals is more or equal to the number of weighted edges between similar opinion individuals. Similarly, the score will always be one if there are edges only between individuals with similar opinions. The final polarization score for the network is then computed by a weighted combination of these two scores (Eq \ref{eq:8}). For block-level understanding of Algorithm \ref{alg:polarization} please refer to Figure \ref{fig:blocks}

\begin{figure*}
  \centering
  \includegraphics[width=\linewidth]{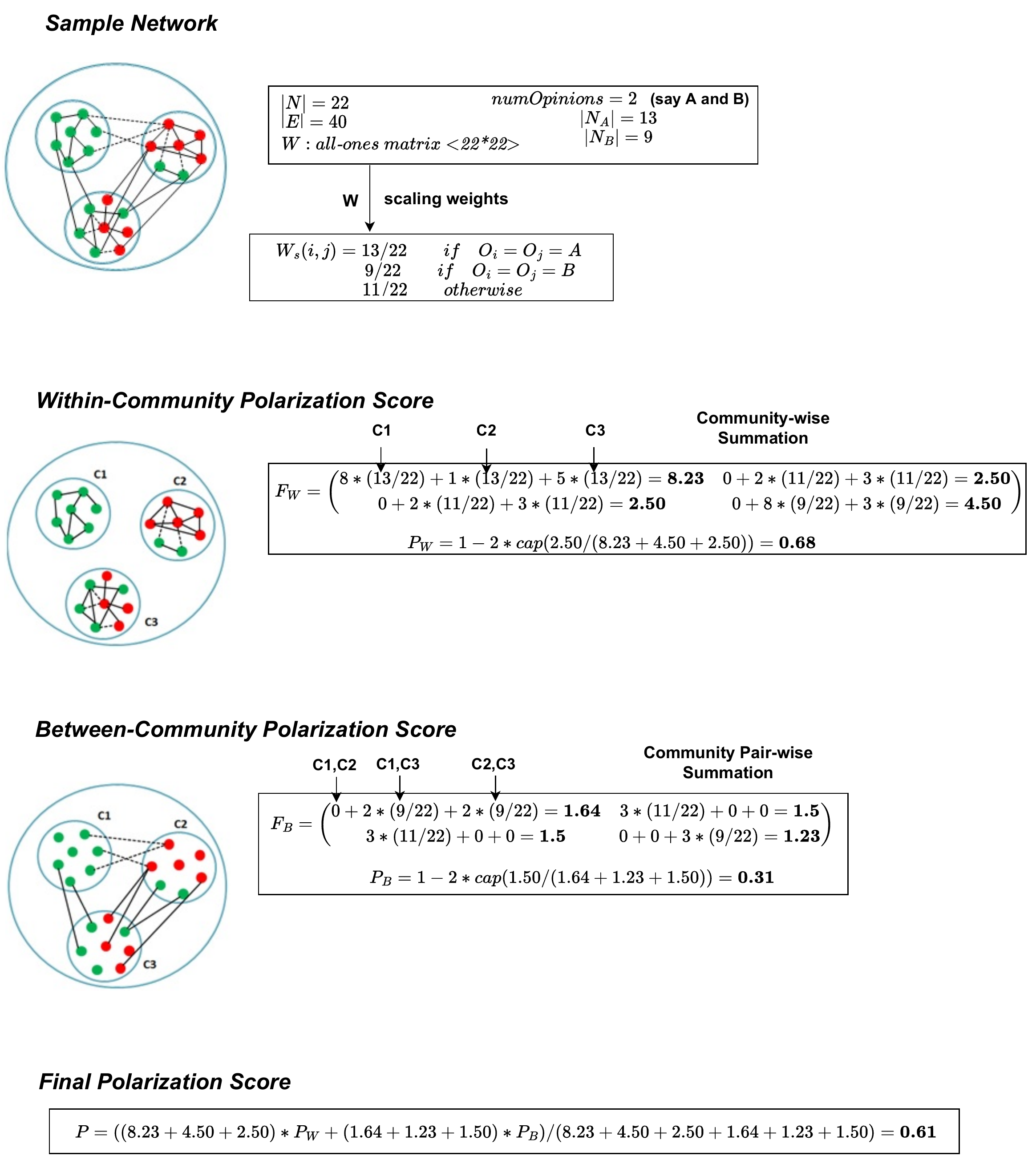}
  \caption{Illustrating our polarization methodology on a sample binary opinion (green: opinion A and red: opinion B) with a solid line edge (connecting similar opinion users) and dashed line (connecting opposing opinion users). The given network comprises three communities ($C1$, $C2$, and $C3$). The matrix $F_B$ considers edges within C1, C2, and C3, whereas matrix $F_B$ considers edges between community pairs C1-C2, C1-C3, and C2-C3.}

  \label{fig:example}
\end{figure*}

To illustrate our proposed method, we have constructed a sample network (Figure \ref{fig:example}). There are a total of 22 nodes and 40 edges in the network. Every node is colored green (13 nodes having opinion A) or red (9 nodes having opinion B). For simplicity, we have assumed all the weights ($W$) to be one. The network is assumed to have three communities, namely $C1$, $C2$, and $C3$. The community $C1$ is considered to be comprised of users having similar opinions (all \say{green}) whereas both the communities $C2$ and $C3$, contains mixed users i.e. users having different opinions (both \say{green} and \say{red} users).  First the weights are scaled using Eq \ref{eq:1} as shown in Figure \ref{fig:example}. Every edge that connects users having opinion A will have a scaled weight of $13/22$, whereas the weight for all the other edges will be $9/22$. Next, the matrices $F_W$ and $F_B$ of size 2*2 (binary opinion scenario) are computed (Eq \ref{eq:2} to 5). For each entry or opinion pair in $F_W$, we computed the community-wise weighted sum of within-community edges (i.e., adding edges from C1, C2, and C3). Similarly, for the matrix $F_B$, we computed the weighted sum of between-community edges (i.e., C1-C2, C1-C3, and C2-C3). Using the matrices $F_W$ and $F_B$, the within-community polarization score i.e. $P_W$ and $P_B$ comes out to be 0.68 (using Eq \ref{eq:6}) and 0.31 (using Eq \ref{eq:7}) respectively. Finally, the polarization score ($P$) for our sample network comes out to be 0.61 (using Eq \ref{eq:8}), which is considered to be an average score in terms of polarization concerning overall communication among the users having conflicting opinions. 

\subsection{Binary Opinion Based Benchmark Networks}
In this section, we compared the performance of our metric with the state-of-the-art techniques for measuring polarization score(\citep{guerra2013measure,garimella2018quantifying,morales2015measuring}). For this, we applied the proposed method to some benchmark networks of different sizes and characteristics. These networks are publicly available and are from diverse domains.

\subsubsection{Karate Club Network}
The network comprises 34 users and 78 edges. It was introduced by Zachary in \citep{zachary1977information} and represented the club's members and administrators as users and the edges as interactions. The graph has two communities due to a conflict between a member and the club administrator. The prior studies have widely used this network to measure polarization (\citep{guerra2013measure,garimella2018quantifying}).

\subsubsection{Political Blogs Network}
A network of U.S. Political Blogs was constructed in \citep{adamic2005political}, where an edge between two blogs indicates the existence of the link from one blog to another. The largest connected component of the network contains 1222 users and 16,714 edges. These users were divided into two ideologies, i.e., liberal and conservative. 

\subsubsection{Political Communication based Retweet Network} \label{sec:political}
Conover et al. \citep{conover2011political} used the Twitter platform to collect tweets using keywords. The users of the retweet network formed from their dataset were labeled using a semi-supervised label propagation method with the help of the publicly expressed opinion of some of the well-known Twitter accounts. The largest connected component, extracted from the retweet network, consists of 18,470 users and 48,053 edges.

As all the above networks are divergent, they become suitable candidates for verifying our polarization metric. 

\subsection{Multi-Opinion Based Synthetic Networks}
All the benchmark networks mentioned in Section \ref{sec:political} assume opinion to be only binary and ignore the possibility of having more than two distinct opinions among users. On the contrary, the method proposed in our study applies to networks with multiple opinions (i.e., for $|O|>=2$). Based on our knowledge, there are no publicly available datasets with more than two distinct opinions; hence we synthetically generated such networks using the political communication-based retweet network mentioned above. We disregarded the original labels of the network and assigned labels in a way to create multiple instances of the same network with different degrees of polarization and a varying number of distinct opinions among users. For this, two parameters were employed: dominance ratio ($domRatio$) and the number of different opinions ($numOpinions$). The first parameter $domRatio$ denotes a proportion of the users having the dominant opinion $X$ within any given community of the network. In contrast, the remaining users in that community will have one of the opinions from the set $O-{X}$ with uniform distribution. Here, the opinion X is randomly selected for each community to ensure that every opinion gets an equal chance to dominate in some community (or communities). On increasing the dominance of any given opinion in a community, it is expected that the network would become more polarized. Similarly, the second parameter $numOpinions$ was used to represent the number of distinct opinions among the users for any given issue. Here, the value of $domRatio$ is kept from 0.3 to 1 (i.e., assuming at least 30\% participation of the given opinion $X$). On the other hand, the number of distinct opinions was kept from 2 to 10 (for simplicity but can be extended beyond the given range also). Thus, both the parameters would take the following set of values:

\begin{gather}
	    domRatio \in\{0.3,0.4,0.5,0.6,0.7,0.8,0.9,1.0\}\\
	    numOpinions \in\{2,3,4,5,6,7,8,9,10\}
	\end{gather}

\subsection{Tri-Opinion Based Retweet Network}\label{sec:trinw}
\begin{figure*}
  \centering
  \includegraphics[width=\linewidth]{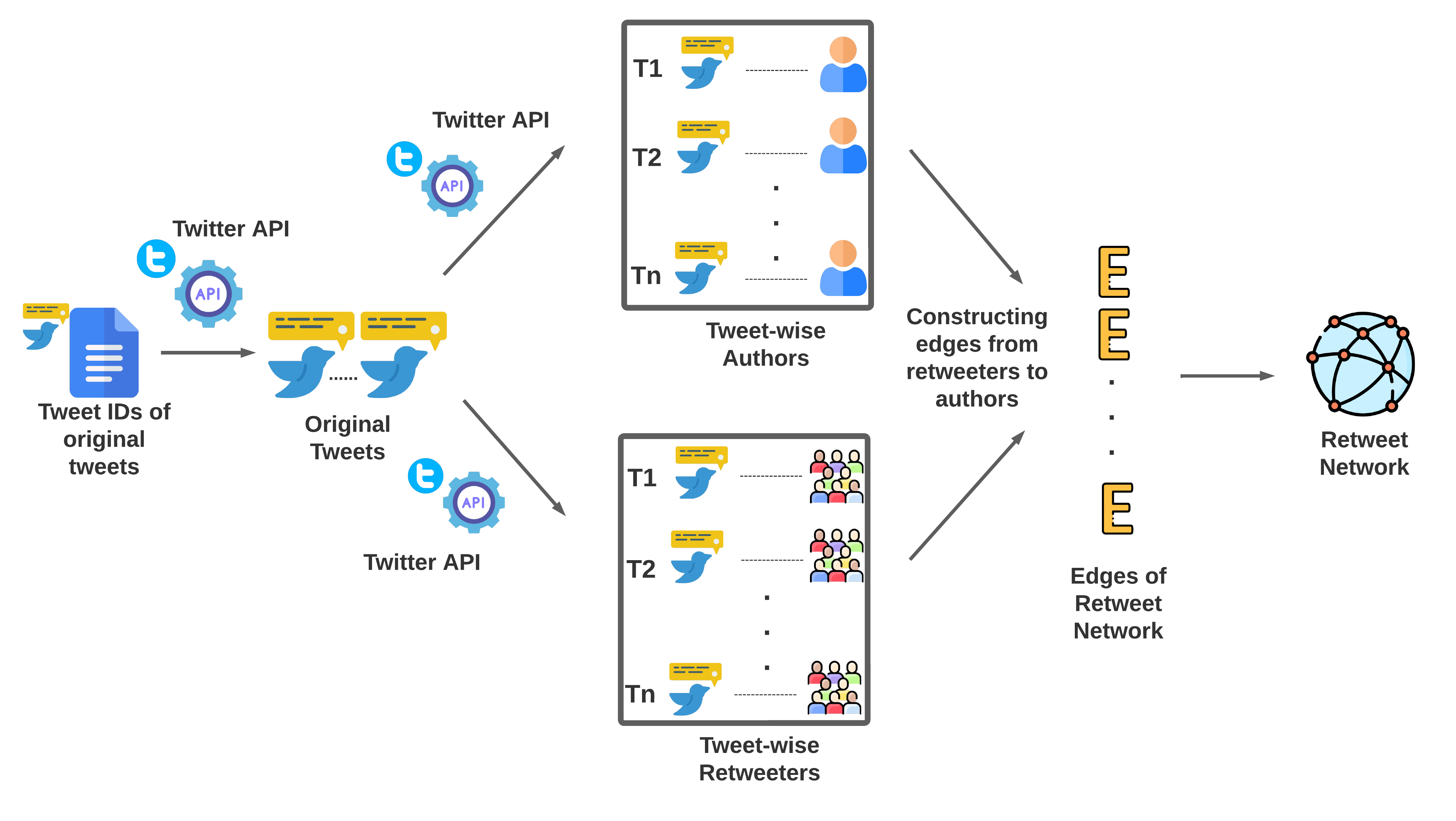}
  \caption{Process of constructing retweet network using the given tweet ids of original tweets regarding COVID-19 associated hydrochloroquine and chloroquine drugs.}

  \label{fig:nwprocess}
\end{figure*}

To quantify polarization using our proposed metric on a retweet network, we used the manually labelled stance based dataset from \citep{mutlu2020stance}. The dataset contained unique identification number (also referred as tweet ids in our case) for 14,374 tweets (does not included any retweets). These tweets were regarding the cure from COVID-19 disease using the drug hydroxychloroquine or chloroquine. The collection period for these tweets were from 1st April 2020 to 30th April 2020. According to \citep{mutlu2020stance}, every tweet contained at least two keywords, first related to COVID-19 (Coronavirus, Corona, COVID-19, Covid19, Sars-cov-2, COVD, Pandemic) and second related to hydroxychloroquine (hydroxychloroquine, chloroquine, and HCQ).Apart from the tweet ids, the dataset also provided the opinion (\say{Favor}, \say{Against} and \say{Neutral}) towards the given issue of each tweet. In order to construct the retweet network of users engaged in discussing the given issue (Figure \ref{fig:nwprocess}), we first leveraged the Twitter API along with the given tweet ids and obtained the details of the users of the given labelled tweets. Out of 14,374 tweets (\say{Favor}: 4685, \say{Against}: 6841, \say{Neutral}: 2848) of the dataset, users for 9294 tweets (\say{Favor}: 3499, \say{Against}: 3776, \say{Neutral}: 2019) were extracted as the other tweets may have been deleted. After obtaining the authors of the original tweets, we again used the Twitter API to extract the users of retweeters of these original tweets. Next, we constructed a weighted retweet network by forming edges from retweeters to the authors of the original tweets. The node in the network denotes the users and there is an edge between every user A and B in the network if either A retweeted B's tweet or B retweeted A's tweet. The weight of an edge would correspond to the total number of times A retweeted B's tweet and B retweeted A's tweets. Later, using the approach given in \citep{grvcar2017stance}, we labelled the opinion of each user based on the opinion of their tweets (assuming that a retweet will have a similar opinion as of its original tweet). First an opinion score ($O_s$) in the range from -1 to 1 was computed for each users using the following equation:-

\begin{equation}
    O_s=(F-A)/(F+A+N)
\end{equation}

Here, $F$, $A$ and $N$ corresponds to the number of favor tweets, against tweets and neutral tweets by each user. Later, we convert these continuous scores for each user $i$ into discrete opinions (i.e. favor, against and neutral) as follows:-

\begin{equation}
    O_i=
    \begin{cases}
        1 & if O_s >0.2\\
        -1 & if O_s<-0.2\\
        0 & otherwise
    \end{cases}
\end{equation} 

The value $O_i$= 1 denotes the opinion of the user $i$ as \say{favor}, -1 for \say{against} and 0 for  \say{Neutral} opinion.

\section{Results and Discussions}
In this section, we have discussed the outcome of implementing our proposed approach on different networks mentioned in the previous section. We first applied the method (Algorithm \ref{alg:polarization}) on three benchmark networks with binary opinions. Next, the proposed multi-opinion approach was tested on 72 synthetically labeled network instances generated from a benchmark network. Inspired by the algorithm efficiency, we employed it on a tri-opinion-based real social network. For all the networks, we have used the Louvain community detection algorithm \citep{blondel2008fast} to obtain the community structure of a given network. As suggested by \citep{aragon2015movement}, any community detection algorithm tends to have a greedy random nature. Therefore, we ran our algorithm 100 times and reported the average scores. This was done for the replication purpose, but even if we executed the algorithm only once, we did not observe any significant change in the scores, thereby ensuring the stability of our algorithm in detecting polarization. The result of implementation on different networks is discussed below:-

\subsection{Binary Opinion Based Benchmark Networks}
On applying the proposed approach to the controversial binary opinion-based benchmark networks, we obtained the following scores:- 0.72 (Karate Network), 0.86 (Blogs Network), and 0.96 (Political Retweet Network). Thus, our approach can capture the overall essence of the level of polarization in all the controversial networks. The relatively higher score for the political retweet network, given by our metric, is in-line with the state-of-the-art methods such as random walk and betweenness-centrality-based measures (Please refer to Table 3 in \citep{garimella2018quantifying}). Moreover, our single metric works on all kinds of networks, unlike the random walk-based measure and the electric dipole moment-based method \citep{garimella2018quantifying}, which produced promising results on large networks but were failed to identify polarization in the smaller network like the karate club network. All these binary opinion-based networks were highly polarized; therefore, to verify the applicability of our approach on a binary labeled network that is not polarized, we applied it to the social relationship network of the fake news and the general news spreaders recently used in \citep{singh2020multidimensional}. Although the labels in the given network are not opinions, the method can be applied to quantify the level of separation between the two groups. The polarization score for the spreaders network comes out to be 0.18, i.e., the two groups are pretty interconnected, which is in line with the claims made in \citep{singh2020multidimensional}. Thus, our method is able to differentiate between the highly polarized network from the non-polarized network with binary opinions (or labels). 

\begin{table*}[t]
 
    \centering
    \caption{Polarization Scores obtained after implementation of Multi-opinion metric on synthetic networks.\\}
    \label{tab:synthetic}
    \begin{tabular}{|c|c|c|c|c|c|c|c|c|}
    \hline
    \multirow{2}{*}{\textbf{$numOpinions$}} & \multicolumn{8}{c|}{\textbf{$domRatio$}}\\
    \cline{2-9}
     & \textbf{0.3} & \textbf{0.4} & \textbf{0.5} & \textbf{0.6} & \textbf{0.7} & \textbf{0.8} & \textbf{0.9} & \textbf{1.0}\\
    \hline
    \textbf{2} & 0.22 & 0.36 & 0.45 & 0.53 & 0.70 & 0.76 & 0.74 & 0.78\\
    \hline
    \textbf{3} & 0.15 & 0.3 & 0.44 & 0.53 & 0.58 & 0.79 & 0.74 & 0.80\\
    \hline
    \textbf{4} & 0.15 & 0.35 & 0.45 & 0.51 & 0.60 & 0.70 & 0.73 & 0.87\\
    \hline
    \textbf{5} & 0.15 & 0.31 & 0.49 & 0.52 & 0.60 & 0.71 & 0.82 & 0.77\\
    \hline
    \textbf{6} & 0.14 & 0.31 & 0.43 & 0.52 & 0.60 & 0.72 & 0.80 & 0.78\\
    \hline
    \textbf{7} & 0.16 & 0.39 & 0.46 & 0.54 & 0.62 & 0.68 & 0.73 & 0.77\\
    \hline
    \textbf{8} & 0.18 & 0.34 & 0.47 & 0.54 & 0.64 & 0.69 & 0.72 & 0.80\\
    \hline
    \textbf{9} & 0.18 & 0.44 & 0.46 & 0.56 & 0.62 & 0.70 & 0.75 & 0.81\\
    \hline
    \textbf{10} & 0.18 & 0.36 & 0.48 & 0.59 & 0.64 & 0.70 & 0.75 & 0.79\\
    \hline

    \end{tabular}
\end{table*}

\subsection{Multi-Opinion Based Synthetic Networks}
As we have seen that the proposed method can quantify polarization on controversial benchmark networks with binary opinions, we are now interested in testing the applicability of our approach on networks of users with multiple opinions and varying levels of polarization. The multi-opinion polarization method is implemented on 72 different networks. The implementations' results are shown in Table \ref{tab:synthetic}. As evident from Table \ref{tab:synthetic}, the scores returned by our method would increase with the increase in the value of the dominance ratio. This behaviour could be attributed to the fact that more the dominance of a single opinion in any given community, more communication between similar opinion users in the network, and hence more polarization among the users in a network. These findings hold for all the given distinct number of opinions among users within the network. Thus, our approach can detect the presence of polarization within the networks with the multi-opinion scenario.

\subsection{Tri-Opinion Based Real Social Network}
\begin{figure*}
  \centering
  \includegraphics[height=10cm,width=10cm]{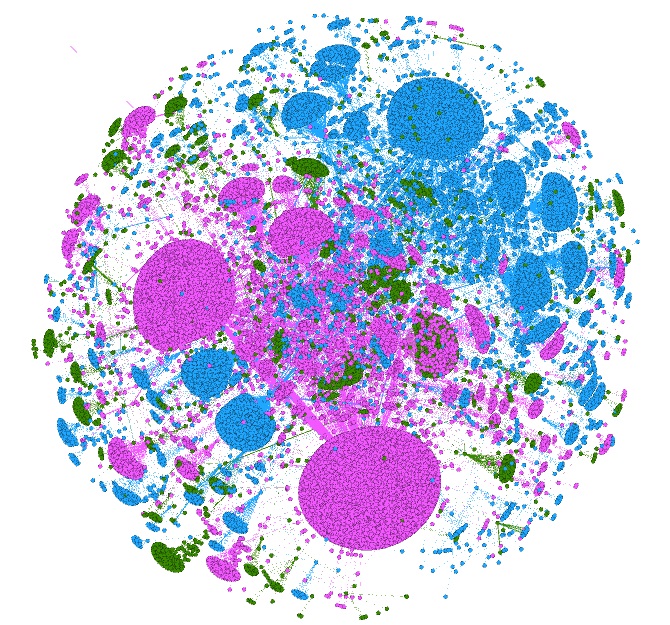}
  \caption{The retweet network represented using Frutcherman Reingold layout \citep{fruchterman1991graph} on Gephi software \citep{bastian2009gephi}.The color of a node represents their opinion towards hydroxychloroquine drug for COVID-19 (blue-Favor, pink-Against and green-Neutral).}

  \label{fig:social}
\end{figure*}

The proposed polarization metric was implemented on a tri-opinion-based retweet network (Users: 37255, Edges: 41668, Communities: 80) of Twitter users (constructed in Section \ref{sec:trinw}) involved in the conversations regarding the usage of hydroxychloroquine for the treatment of COVID-19 (Figure \ref{fig:social}. Our algorithm returned a score of 0.93 for the given retweet network. This high score indicates that Twitter users had polarized opinions about hydroxychloroquine or chloroquine drugs. These findings are contrary to another study \citep{singh2022twitter}, which focused on users' sentiment toward COVID-19-associated Mucormycosis. The retweet network in that study was instead found to be non-polarized. Based on the prior studies, there could be multiple reasons for polarized opinions towards the usage of the drug. One of the reasons could be the absence of clear scientific evidence that drugs like hydroxychloroquine or chloroquine effectively cure patients suffering from COVID-19 disease \citep{lenzer2020covid}. Another reason for could be the severe side effects associated with these drugs \citep{meyerowitz2020rethinking}. The third reason could be the the infusion of politics into the matter, as highlighted in \citep{saag2020misguided}. Future research is required in this regard to determine the potential cause(s). Overall, it can be stated that the given issue bifurcated the users on Twitter based on opinion, where one was in support of the drug, another was against the usage, and the remaining users were neither in favor nor opposed the use of the drug.

\section{Conclusion}
In this study, we have proposed a method for quantifying polarization on online social networks. The proposed method is applicable for a multi-opinion scenario, such as controversies involving multiple sides. The polarization score returned by our method is the weighted sum of two different polarization scores: within-community and between-community. The within-community scores quantify polarization among users present within the same community. On the other hand, the between-community score quantifies polarization among users belonging to different communities. The intuition behind both the scores is that the level of opinion polarization in any network is correlated with the proportion of communication or connections between users of opposing opinions. We verified the applicability of our method on networks with different degrees of polarization and the varying number of opinions. Our approach also suggested a high polarization among users towards the usage of hydroxychloroquine and chloroquine drugs for the prevention of COVID-19 disease. This high score indicates uncertainty among the users regarding the benefits of the drugs. In the future, we would like to study polarization as a dynamic process using our proposed metric on different kinds of social networks.

\bibliographystyle{unsrtnat}
\bibliography{pol}  

\begin{thebibliography}{43}
\providecommand{\natexlab}[1]{#1}
\providecommand{\url}[1]{\texttt{#1}}
\expandafter\ifx\csname urlstyle\endcsname\relax
  \providecommand{\doi}[1]{doi: #1}\else
  \providecommand{\doi}{doi: \begingroup \urlstyle{rm}\Url}\fi

\bibitem[Valenzuela(2013)]{valenzuela2013unpacking}
Sebasti{\'a}n Valenzuela.
\newblock Unpacking the use of social media for protest behavior: The roles of
  information, opinion expression, and activism.
\newblock \emph{American behavioral scientist}, 57\penalty0 (7):\penalty0
  920--942, 2013.

\bibitem[Conover et~al.(2011)Conover, Ratkiewicz, Francisco, Gon{\c{c}}alves,
  Menczer, and Flammini]{conover2011political}
Michael~D Conover, Jacob Ratkiewicz, Matthew~R Francisco, Bruno
  Gon{\c{c}}alves, Filippo Menczer, and Alessandro Flammini.
\newblock Political polarization on twitter.
\newblock \emph{Icwsm}, 133\penalty0 (26):\penalty0 89--96, 2011.

\bibitem[Bossetta et~al.(2018)Bossetta, Segesten, and
  Trenz]{bossetta2018political}
Michael Bossetta, Anamaria~Dutceac Segesten, and Hans-J{\"o}rg Trenz.
\newblock Political participation on facebook during brexit: Does user
  engagement on media pages stimulate engagement with campaigns?
\newblock \emph{Journal of Language and Politics}, 17\penalty0 (2):\penalty0
  173--194, 2018.

\bibitem[Miller(2019)]{miller2019americans}
Steven~V Miller.
\newblock What americans think about gun control: Evidence from the general
  social survey, 1972--2016.
\newblock \emph{Social Science Quarterly}, 100\penalty0 (1):\penalty0 272--288,
  2019.

\bibitem[Mouw and Sobel(2001)]{mouw2001culture}
Ted Mouw and Michael~E Sobel.
\newblock Culture wars and opinion polarization: the case of abortion.
\newblock \emph{American Journal of Sociology}, 106\penalty0 (4):\penalty0
  913--943, 2001.

\bibitem[Lewis et~al.(2017)Lewis, Flores, Haider-Markel, Miller, Tadlock, and
  Taylor]{lewis2017degrees}
Daniel~C Lewis, Andrew~R Flores, Donald~P Haider-Markel, Patrick~R Miller,
  Barry~L Tadlock, and Jami~K Taylor.
\newblock Degrees of acceptance: Variation in public attitudes toward segments
  of the lgbt community.
\newblock \emph{Political Research Quarterly}, 70\penalty0 (4):\penalty0
  861--875, 2017.

\bibitem[Goldberg and Savenije(2018)]{goldberg2018teaching}
Tsafrir Goldberg and Geerte~M Savenije.
\newblock Teaching controversial historical issues.
\newblock \emph{The Wiley international handbook of history teaching and
  learning}, pages 503--526, 2018.

\bibitem[Gallagher et~al.(2018)Gallagher, Reagan, Danforth, and
  Dodds]{gallagher2018divergent}
Ryan~J Gallagher, Andrew~J Reagan, Christopher~M Danforth, and Peter~Sheridan
  Dodds.
\newblock Divergent discourse between protests and counter-protests:\#
  blacklivesmatter and\# alllivesmatter.
\newblock \emph{PloS one}, 13\penalty0 (4):\penalty0 e0195644, 2018.

\bibitem[Yuan et~al.(2019)Yuan, Schuchard, and Crooks]{yuan2019examining}
Xiaoyi Yuan, Ross~J Schuchard, and Andrew~T Crooks.
\newblock Examining emergent communities and social bots within the polarized
  online vaccination debate in twitter.
\newblock \emph{Social media+ society}, 5\penalty0 (3):\penalty0
  2056305119865465, 2019.

\bibitem[DiMaggio et~al.(1996)DiMaggio, Evans, and Bryson]{dimaggio1996have}
Paul DiMaggio, John Evans, and Bethany Bryson.
\newblock Have american's social attitudes become more polarized?
\newblock \emph{American journal of Sociology}, 102\penalty0 (3):\penalty0
  690--755, 1996.

\bibitem[Gay et~al.(1996)Gay, Ellison, and Powers]{gay1996search}
David~A Gay, Christopher~G Ellison, and Daniel~A Powers.
\newblock In search of denominational subcultures: Religious affiliation and"
  pro-family" issues revisited.
\newblock \emph{Review of Religious Research}, pages 3--17, 1996.

\bibitem[Banisch and Olbrich(2019)]{banisch2019opinion}
Sven Banisch and Eckehard Olbrich.
\newblock Opinion polarization by learning from social feedback.
\newblock \emph{The Journal of Mathematical Sociology}, 43\penalty0
  (2):\penalty0 76--103, 2019.

\bibitem[Koudenburg et~al.(2021)Koudenburg, Kiers, and
  Kashima]{koudenburg2021new}
Namkje Koudenburg, Henk~AL Kiers, and Yoshihisa Kashima.
\newblock A new opinion polarization index developed by integrating expert
  judgments.
\newblock \emph{Frontiers in psychology}, page 4575, 2021.

\bibitem[Morales et~al.(2015)Morales, Borondo, Losada, and
  Benito]{morales2015measuring}
Alfredo~Jose Morales, Javier Borondo, Juan~Carlos Losada, and Rosa~M Benito.
\newblock Measuring political polarization: Twitter shows the two sides of
  venezuela.
\newblock \emph{Chaos: An Interdisciplinary Journal of Nonlinear Science},
  25\penalty0 (3):\penalty0 033114, 2015.

\bibitem[Garimella et~al.(2018{\natexlab{a}})Garimella, De~Francisci~Morales,
  Gionis, and Mathioudakis]{garimella2018political}
Kiran Garimella, Gianmarco De~Francisci~Morales, Aristides Gionis, and Michael
  Mathioudakis.
\newblock Political discourse on social media: Echo chambers, gatekeepers, and
  the price of bipartisanship.
\newblock In \emph{Proceedings of the 2018 World Wide Web Conference}, pages
  913--922, 2018{\natexlab{a}}.

\bibitem[Newman(2006)]{newman2006modularity}
Mark~EJ Newman.
\newblock Modularity and community structure in networks.
\newblock \emph{Proceedings of the national academy of sciences}, 103\penalty0
  (23):\penalty0 8577--8582, 2006.

\bibitem[Guerra et~al.(2013)Guerra, Meira~Jr, Cardie, and
  Kleinberg]{guerra2013measure}
Pedro Henrique~Calais Guerra, Wagner Meira~Jr, Claire Cardie, and Robert
  Kleinberg.
\newblock A measure of polarization on social media networks based on community
  boundaries.
\newblock In \emph{ICWSM}, 2013.

\bibitem[Garimella and Weber(2017)]{garimella2017long}
Kiran Garimella and Ingmar Weber.
\newblock A long-term analysis of polarization on twitter.
\newblock \emph{arXiv preprint arXiv:1703.02769}, 2017.

\bibitem[Garimella et~al.(2018{\natexlab{b}})Garimella, Morales, Gionis, and
  Mathioudakis]{garimella2018quantifying}
Kiran Garimella, Gianmarco De~Francisci Morales, Aristides Gionis, and Michael
  Mathioudakis.
\newblock Quantifying controversy on social media.
\newblock \emph{ACM Transactions on Social Computing}, 1\penalty0 (1):\penalty0
  1--27, 2018{\natexlab{b}}.

\bibitem[Darwish(2019)]{darwish2019quantifying}
Kareem Darwish.
\newblock Quantifying polarization on twitter: The kavanaugh nomination.
\newblock In \emph{International Conference on Social Informatics}, pages
  188--201. Springer, 2019.

\bibitem[Girvan and Newman(2002)]{girvan2002community}
Michelle Girvan and Mark~EJ Newman.
\newblock Community structure in social and biological networks.
\newblock \emph{Proceedings of the national academy of sciences}, 99\penalty0
  (12):\penalty0 7821--7826, 2002.

\bibitem[Radicchi et~al.(2004)Radicchi, Castellano, Cecconi, Loreto, and
  Parisi]{radicchi2004defining}
Filippo Radicchi, Claudio Castellano, Federico Cecconi, Vittorio Loreto, and
  Domenico Parisi.
\newblock Defining and identifying communities in networks.
\newblock \emph{Proceedings of the national academy of sciences}, 101\penalty0
  (9):\penalty0 2658--2663, 2004.

\bibitem[Labatut et~al.(2014)Labatut, Dugu{\'e}, and
  Perez]{labatut2014identifying}
Vincent Labatut, Nicolas Dugu{\'e}, and Anthony Perez.
\newblock Identifying the community roles of social capitalists in the twitter
  network.
\newblock In \emph{2014 IEEE/ACM International Conference on Advances in Social
  Networks Analysis and Mining (ASONAM 2014)}, pages 371--374. IEEE, 2014.

\bibitem[Singh et~al.(2022{\natexlab{a}})Singh, Iyengar, and
  Kaur]{singh2022mining}
Maneet Singh, SRS Iyengar, and Rishemjit Kaur.
\newblock Mining social networks for dissemination of fake news using
  continuous opinion-based hybrid model.
\newblock In \emph{International Conference on Advanced Data Mining and
  Applications}, pages 217--228. Springer, 2022{\natexlab{a}}.

\bibitem[ani(2019)]{ani_news2019}
Aap only supported centre on article 370, never backed idea of j-k as ut:
  Sanjay singh, Aug 2019.
\newblock URL \url{https://urlcc.cc/83a2r}.

\bibitem[Mutlu et~al.(2020)Mutlu, Oghaz, Jasser, Tutunculer, Rajabi, Tayebi,
  Ozmen, and Garibay]{mutlu2020stance}
Ece~C Mutlu, Toktam Oghaz, Jasser Jasser, Ege Tutunculer, Amirarsalan Rajabi,
  Aida Tayebi, Ozlem Ozmen, and Ivan Garibay.
\newblock A stance data set on polarized conversations on twitter about the
  efficacy of hydroxychloroquine as a treatment for covid-19.
\newblock \emph{Data in brief}, 33:\penalty0 106401, 2020.

\bibitem[Sinha and Balayla(2020)]{sinha2020hydroxychloroquine}
Neeraj Sinha and Galit Balayla.
\newblock Hydroxychloroquine and covid-19.
\newblock \emph{Postgraduate medical journal}, 96\penalty0 (1139):\penalty0
  550--555, 2020.

\bibitem[Piszczatoski and Powell(2020)]{piszczatoski2020emergency}
Christopher~R Piszczatoski and Jason Powell.
\newblock Emergency authorization of chloroquine and hydroxychloroquine for
  treatment of covid-19.
\newblock \emph{Annals of Pharmacotherapy}, 54\penalty0 (8):\penalty0 827--831,
  2020.

\bibitem[Li et~al.(2020)Li, Wang, Agostinis, Rabson, Melino, Carafoli, Shi, and
  Sun]{li2020hydroxychloroquine}
Xing Li, Ying Wang, Patrizia Agostinis, Arnold Rabson, Gerry Melino, Ernesto
  Carafoli, Yufang Shi, and Erwei Sun.
\newblock Is hydroxychloroquine beneficial for covid-19 patients?
\newblock \emph{Cell death \& disease}, 11\penalty0 (7):\penalty0 1--6, 2020.

\bibitem[Manivannan et~al.(2021)Manivannan, Karthikeyan, Moorthy, and
  Chaturvedi]{manivannan2021rise}
Elangovan Manivannan, Chandrabose Karthikeyan, NS~Moorthy, and Subash~Chandra
  Chaturvedi.
\newblock The rise and fall of chloroquine/hydroxychloroquine as compassionate
  therapy of covid-19.
\newblock \emph{Frontiers in Pharmacology}, 12:\penalty0 1057, 2021.

\bibitem[Blondel et~al.(2008)Blondel, Guillaume, Lambiotte, and
  Lefebvre]{blondel2008fast}
Vincent~D Blondel, Jean-Loup Guillaume, Renaud Lambiotte, and Etienne Lefebvre.
\newblock Fast unfolding of communities in large networks.
\newblock \emph{Journal of statistical mechanics: theory and experiment},
  2008\penalty0 (10):\penalty0 P10008, 2008.

\bibitem[Olivares et~al.(2019)Olivares, C{\'a}rdenas, Losada, and
  Borondo]{olivares2019opinion}
Gast{\'o}n Olivares, Juan~Pablo C{\'a}rdenas, Juan~Carlos Losada, and Javier
  Borondo.
\newblock Opinion polarization during a dichotomous electoral process.
\newblock \emph{Complexity}, 2019, 2019.

\bibitem[Zachary(1977)]{zachary1977information}
Wayne~W Zachary.
\newblock An information flow model for conflict and fission in small groups.
\newblock \emph{Journal of anthropological research}, 33\penalty0 (4):\penalty0
  452--473, 1977.

\bibitem[Adamic and Glance(2005)]{adamic2005political}
Lada~A Adamic and Natalie Glance.
\newblock The political blogosphere and the 2004 us election: divided they
  blog.
\newblock In \emph{Proceedings of the 3rd international workshop on Link
  discovery}, pages 36--43, 2005.

\bibitem[Gr{\v{c}}ar et~al.(2017)Gr{\v{c}}ar, Cherepnalkoski, Mozeti{\v{c}},
  and Kralj~Novak]{grvcar2017stance}
Miha Gr{\v{c}}ar, Darko Cherepnalkoski, Igor Mozeti{\v{c}}, and Petra
  Kralj~Novak.
\newblock Stance and influence of twitter users regarding the brexit
  referendum.
\newblock \emph{Computational social networks}, 4\penalty0 (1):\penalty0 1--25,
  2017.

\bibitem[Arag{\'o}n et~al.(2015)Arag{\'o}n, Volkovich, Laniado, and
  Kaltenbrunner]{aragon2015movement}
Pablo Arag{\'o}n, Yana Volkovich, David Laniado, and Andreas Kaltenbrunner.
\newblock When a movement becomes a party: The 2015 barcelona city council
  election.
\newblock \emph{arXiv preprint arXiv:1507.08599}, 2015.

\bibitem[Singh et~al.(2020)Singh, Kaur, and Iyengar]{singh2020multidimensional}
Maneet Singh, Rishemjit Kaur, and SRS Iyengar.
\newblock Multidimensional analysis of fake news spreaders on twitter.
\newblock In \emph{International Conference on Computational Data and Social
  Networks}, pages 354--365. Springer, 2020.

\bibitem[Singh et~al.(2022{\natexlab{b}})Singh, Dhillon, Ichhpujani, Iyengar,
  and Kaur]{singh2022twitter}
Maneet Singh, Hennaav~Kaur Dhillon, Parul Ichhpujani, Sudarshan Iyengar, and
  Rishemjit Kaur.
\newblock Twitter sentiment analysis for covid-19 associated mucormycosis.
\newblock \emph{Indian Journal of Ophthalmology}, 70\penalty0 (5):\penalty0
  1773--1779, 2022{\natexlab{b}}.

\bibitem[Lenzer(2020)]{lenzer2020covid}
Jeanne Lenzer.
\newblock Covid-19: Us gives emergency approval to hydroxychloroquine despite
  lack of evidence.
\newblock \emph{bmj}, 369\penalty0 (10.1136), 2020.

\bibitem[Meyerowitz et~al.(2020)Meyerowitz, Vannier, Friesen, Schoenfeld,
  Gelfand, Callahan, Kim, Reeves, and Poznansky]{meyerowitz2020rethinking}
Eric~A Meyerowitz, Augustin~GL Vannier, Morgan~GN Friesen, Sara Schoenfeld,
  Jeffrey~A Gelfand, Michael~V Callahan, Arthur~Y Kim, Patrick~M Reeves, and
  Mark~C Poznansky.
\newblock Rethinking the role of hydroxychloroquine in the treatment of
  covid-19.
\newblock \emph{The FASEB Journal}, 34\penalty0 (5):\penalty0 6027--6037, 2020.

\bibitem[Saag(2020)]{saag2020misguided}
Michael~S Saag.
\newblock Misguided use of hydroxychloroquine for covid-19: the infusion of
  politics into science.
\newblock \emph{Jama}, 324\penalty0 (21):\penalty0 2161--2162, 2020.

\bibitem[Fruchterman and Reingold(1991)]{fruchterman1991graph}
Thomas~MJ Fruchterman and Edward~M Reingold.
\newblock Graph drawing by force-directed placement.
\newblock \emph{Software: Practice and experience}, 21\penalty0 (11):\penalty0
  1129--1164, 1991.

\bibitem[Bastian et~al.(2009)Bastian, Heymann, and Jacomy]{bastian2009gephi}
Mathieu Bastian, Sebastien Heymann, and Mathieu Jacomy.
\newblock Gephi: an open source software for exploring and manipulating
  networks.
\newblock In \emph{Proceedings of the International AAAI Conference on Web and
  Social Media}, volume~3, 2009.

\end{thebibliography}

\end{document}